\documentclass[12pt]{article}

\newcommand{\x}{arXiv:}
\newcommand{\m}{\mathrm}
\newcommand{\be}{\begin{equation}}
\newcommand{\ee}{\end{equation}}
\newcommand{\ba}{\begin{eqnarray}}
\newcommand{\ea}{\end{eqnarray}}
\newcommand{\dif}{\mathrm{d}}

\usepackage{graphicx}
\usepackage{amssymb}
\usepackage{amsmath}
\usepackage[T1]{fontenc} 
\usepackage[ansinew]{inputenc} 
\usepackage[nosort]{cite}
\newcommand{\inbar}{\vrule height1.57ex width.4pt depth0pt}
\newcommand{\SW}{\relax{\hbox{$\ \inbar\kern-.285em{\rm S}$}}}

\begin{document}
\thispagestyle{empty}
\begin{center}

\null \vskip-1truecm \vskip2truecm

{\Large{\bf \textsf{Inverse Magnetic/Shear Catalysis}}}

{\Large{\bf \textsf{}}}

{\large{\bf \textsf{}}}

{\large{\bf \textsf{}}}

\vskip1truecm

{\large \textsf{Brett McInnes
}}

\vskip0.1truecm

\textsf{\\ National
  University of Singapore}
  \vskip1.2truecm
\textsf{email: matmcinn@nus.edu.sg}\\

\end{center}
\vskip1truecm \centerline{\textsf{ABSTRACT}} \baselineskip=15pt

\medskip
It is well known that very large magnetic fields are generated when the Quark-Gluon Plasma is formed during peripheral heavy-ion collisions. Lattice, holographic, and other studies strongly suggest that these fields may, for observationally relevant field values, induce ``inverse magnetic catalysis'', signalled by a lowering of the critical temperature for the chiral/deconfinement transition. The theoretical basis of this effect has recently attracted much attention; yet so far these investigations have not included another, equally dramatic consequence of the peripheral collision geometry: the QGP acquires a large angular momentum vector, parallel to the magnetic field. Here we use holographic techniques to argue that the angular momentum can also, independently, have an effect on transition temperatures, and we obtain a rough estimate of the relative effects of the presence of both a magnetic field and an angular momentum density. We find that the shearing angular momentum reinforces the effect of the magnetic field at low values of the baryonic chemical potential, but that it can actually decrease that effect at high chemical potentials.

\newpage
\addtocounter{section}{1}
\section* {\large{\textsf{1. The QGP Subjected to Magnetic Fields and Shear}}}
The ability of current facilities \cite{kn:STAR,kn:ALICE,kn:BEAM,kn:armesto} to investigate deconfined quark matter in the form of a \emph{Quark-Gluon Plasma} (QGP), resulting from collisions of heavy ions, represents a major step forward, and it is vitally important that this form of matter be understood theoretically \cite{kn:vuor}. However, collisions of these extended and complex objects produce a QGP which is not simple. In particular, while some collisions are \emph{central}, others are \emph{peripheral} (``off-centre''), and this leads to several complications.

One such complication, which has received a great deal of attention, is that the QGP produced in a peripheral collision can be subject to an extremely large magnetic field \cite{kn:skokov,kn:tuchin,kn:magnet}. The magnetic fields involved are so large that they can affect parameters that would otherwise be exclusively in the domain of the strong interaction, and this has led to predictions of various novel phenomena in the context of what is effectively a new branch of quantum chromodynamics \cite{kn:review} (``magnetochromodynamics'').

In particular, there are very general theoretical arguments to the effect that magnetic fields of this order might affect the chiral/deconfinement transition, and particularly the (pseudo-)critical temperature at which the latter occurs (at zero baryonic chemical potential). It is thought that this could be due to a reduction of the ``effective dimensionality'' imposed by the magnetic field. This idea will be the basis of our discussions in this work.

Until recently, it was thought that the effect of strong magnetic fields would be to ``catalyse'' the chiral transition, resulting in a higher transition temperature\footnote{In principle, the chiral and deconfinement transitions are of course different, and it is conceivable that magnetic fields might catalyse the chiral transition while yet driving down the deconfinement temperature. This has been discussed from a holographic point of view in \cite{kn:dudal}. Here we confine our attention to the standard picture in which the two transitions always occur in the same way.}: this is ``magnetic catalysis''. However, lattice computations \cite{kn:bali} suggest the opposite effect on the transition temperature: this is \emph{inverse} magnetic catalysis ---$\,$ see for example \cite{kn:ferrer,kn:ayala1} for recent discussions of this effect, and \cite{kn:naylor,kn:fraga} for reviews. This might mean that the transition occurs at a lower temperature in a plasma produced by a peripheral collision than in a plasma associated with a central collision in the same beam; whether such an effect is actually observable remains to be seen.

Still more recently, it has been argued (for references, see again the reviews \cite{kn:naylor,kn:fraga}) that analogous phenomena may also occur at non-zero values of the baryonic chemical potential, in particular, near to the critical point which is generally thought to exist in the quark matter phase diagram \cite{kn:mohanty,kn:satz}. We shall consider the holography of this case too; it will be important for experiments in the near future.

The theoretical basis of inverse magnetic catalysis is not well understood, and constructing a convincing theory is a major objective of current theoretical research (see for example \cite{kn:ayala2} for one set of suggestions). At this point, it is important to subject any theory of this effect to tests under a wide variety of conditions\footnote{The simplest, indeed perhaps overly simple models, are the ``magnetic bag models''; for a discussion of their drawbacks and uses, see \cite{kn:fraga}.}. For example, recent lattice results \cite{kn:gergely} indicate that inverse catalysis holds even at ultra-high fields (well over 3 GeV$^2$, and perhaps even indefinitely), though there are other theoretical arguments to the effect that it might be replaced by magnetic \emph{catalysis} at still higher field values \cite{kn:shov}. This is of great interest for testing theoretical proposals, even though such high fields do not occur in collisions producing temperatures near to the transition temperatures.

Inverse magnetic catalysis is associated with the internal motion of the QGP when it is produced by a peripheral collision. But this internal motion is also associated with another phenomenon: \emph{the plasma inevitably acquires a very large angular momentum} (per unit energy). This angular momentum can be associated with either rotation \cite{kn:KelvinHelm,kn:viscous,kn:csernairecent1,kn:csernairecent2} (see \cite{kn:nagy} for the most recent developments) or shear \cite{kn:liang,kn:bec,kn:huang}; the latter will be our concern here. The shearing motion is represented by an angular momentum vector parallel to the magnetic field ---$\,$ that is, perpendicular to the reaction plane (the space conventionally used to represent a typical section through the plasma). We stress that these two vectors are inextricably related: the angular momentum vector has hitherto been neglected in theoretical studies of inverse catalysis not because it is absent, but because it does not seem to be relevant.

One approach to the study of the QGP involves the use of the gauge-gravity duality \cite{kn:youngman,kn:gubser,kn:janik}, and such methods have indeed been used to study such issues as the effect of magnetic fields on equilibration \cite{kn:yaffe,kn:mamoo} as well as inverse magnetic catalysis \cite{kn:mamo,kn:rouge}. Holographic methods can be used to study magnetic fields because a magnetic charge on the bulk black hole has an effect which persists to infinity \cite{kn:hartkov}. However, in the asymptotically AdS case, a black hole with angular momentum distorts the bulk spacetime in a characteristic way (through frame-dragging), and this effect \emph{also} persists to infinity. In the case of a black hole having an event horizon with spherical topology, the effect at infinity is to induce rotation \cite{kn:sonner,kn:schalm}; but, in the case of a non-spherical event horizon \cite{kn:lemmo}, black hole angular momentum \cite{kn:klemm} induces a shearing velocity profile in the space representing the two-dimensional reaction plane\footnote{Thus, the \emph{spacetime} at infinity is three-dimensional, described by the conventional collision physics coordinates $(x, z, t)$ (with $z$ representing the collision axis), and the bulk is four-dimensional here.} at infinity \cite{kn:75,kn:shear,kn:79}.

The effect of angular momentum on the black hole is in many ways similar to that of magnetic charge (though there are some important differences). Because of this, and because both have a direct influence on the geometry at infinity, the gauge-gravity duality leads us to suspect that the shearing motion in the QGP \emph{will have a similar effect to that of a magnetic field}. In particular, the shear should either reinforce or oppose the effect of the magnetic field on the (pseudo)critical temperature.

Our attitude here is that the gauge-gravity duality is not to be trusted to produce precise numerical results: however, it may well be a reliable guide to trends. Our main objective is to determine these trends. We will argue that holography implies that the \emph{combined} effect of magnetic fields and angular momentum is indeed to reduce the transition temperature: holography, including angular momentum, predicts inverse catalysis\footnote{We find no evidence for an increase in the transition temperature at any value of the magnetic field, in agreement with \cite{kn:gergely}.}. Since the two effects are, as explained earlier, inextricably associated, we will speak of \emph{inverse magnetic/shear catalysis}.

Granted that shear contributes to inverse catalysis, the next question is whether it reinforces or \emph{opposes} the effect of the magnetic field. We find that the situation depends critically on whether the baryonic chemical potential $\mu_B$ is zero. If it is, then shear always reinforces inverse magnetic catalysis, though the effect, for reasonable parameter values, is small. When $\mu_B \neq 0$, however, the situation is more complex and interesting: for relatively low angular momenta, \emph{shear actually works to reduce the effects of the magnetic field} (though it never succeeds in overcoming it completely); it only begins to reinforce it when the angular momentum is extremely large. The hope is that the complex pattern just described will present a useful challenge to theories of the underlying mechanisms of inverse catalysis.

We begin with a description of the bulk dual to a boundary field theory subject to both an intense magnetic field and a strong shearing motion. This may well be of general interest, because this is the \emph{generic} situation produced by a collision of heavy ions, and also because the black holes we shall consider (belonging to the Pleba\'nski--Demia\'nski family of metrics) are the most complex known \emph{exact} black hole solutions of the AdS Einstein-Maxwell equations incorporating angular momentum\footnote{Strictly speaking, there is a still more complex family arising from the Pleba\'nski--Demia\'nski ``acceleration'' parameter. However, we argued in \cite{kn:shear} that Pleba\'nski--Demia\'nski solutions with non-vanishing acceleration parameter are not suited to a holographic description of the shearing plasma.}. Holographic investigations beyond this point will involve either ad hoc deformations of the bulk geometry (which may then fail to satisfy the AdS Einstein-Maxwell equations, with potentially serious consequences for the internal consistency of the model) or highly sophisticated numerical techniques. (Numerical investigations of holographic duals of off-centre collisions have indeed recently begun: see \cite{kn:chesyaf}.)

We will be interested in a boundary field theory which models a plasma described by a very basic set of physical parameters: the temperature $T$, mean magnetic field $B_m$, and specific angular momentum $a$ (that is, angular momentum per unit mass or energy). The objective is to understand how $T$ changes as $B_m$ and $a$ are increased; this increase corresponds to considering various possible impact parameters, since the amount of angular momentum transferred to the plasma varies from event to event, reaching a peak for a certain optimal impact parameter \cite{kn:bec}. That is, we wish to study the way $T$ varies for various collisions in a given beam. The beam itself will be regarded as setting initial conditions for the plasma, through an average specific entropy (entropy per unit energy) $\varsigma_S$, and a specific baryonic chemical potential (chemical potential per unit temperature) $\varsigma_B \equiv \mu_B/T$ (which is approximately zero at the RHIC and LHC experiments, but not in experiments planned for the near future).

We also need to prescribe a precise velocity profile describing the shearing motion in the QGP. This is specified by a function $v(x)$, which gives the velocity (in natural units, so it is dimensionless) of the QGP as a function of the coordinate $x$, transverse to the collision axis $z$. The form of this function is determined by the impact parameter and the distribution of the nucleons in the projectile nuclei: see \cite{kn:bec} for a clear discussion. In general, it rises from zero along the symmetry axis to some maximum\footnote{We require that $V$ satisfy $V \leq 1$. The motivation for this is obvious on the boundary; the dual interpretation is that it arises from the need to avoid a superradiant instability in the bulk: see \cite{kn:green}.} $V$ at the boundary of the collision zone; $V$ is determined by the time since the initial impact and other parameters \cite{kn:KelvinHelm}. While various general shapes consistent with hydrodynamic stability are possible \cite{kn:shear}, for definiteness we shall consider a function\footnote{The precise functional form will be given below; in this particular example one has $V \approx 0.4$.} of the form shown in Figure 1. This shape is distinguished simply by the fact that it is both fairly realistic (see for example Figure 3 in \cite{kn:huang}), and that it is one of only two for which the dual bulk geometry is known exactly. (The other shape is discussed in \cite{kn:shear}: it is only suited to a discussion of the plasma close to the symmetry axis.)
\begin{figure}[!h]
\centering
\includegraphics[width=0.55\textwidth]{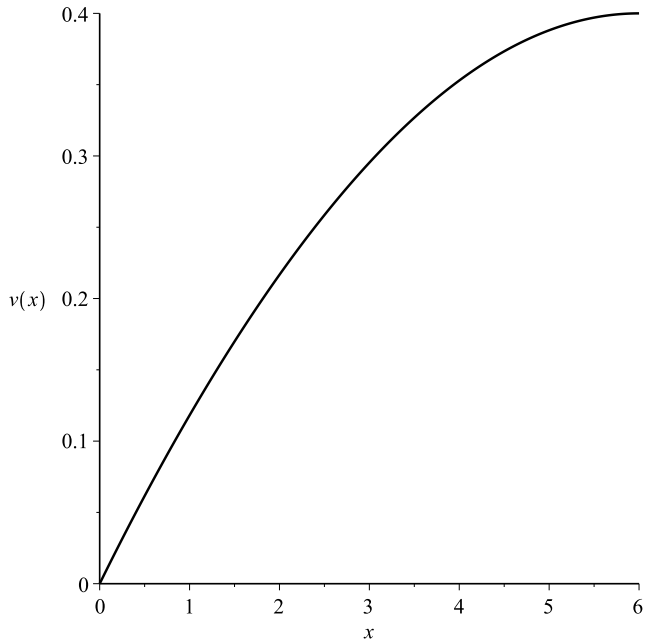}
\caption{Example of a shearing velocity profile.}
\end{figure}

In order to proceed, we therefore need a gauge-gravity dual model, and a ``dictionary'' that converts all of these parameters and functions to quantities describing an asymptotically AdS black hole. We begin with a description of the dual black hole.

\addtocounter{section}{1}
\section* {\large{\textsf{2. The Bulk Black Hole and its Parameters}}}
A black hole, with topologically planar event horizon ---$\,$ the boundary is in fact globally conformally flat ---$\,$ having a metric inducing a shearing motion at infinity, was first given by Klemm et al. \cite{kn:klemm}; we have called it the ``KMV$_0$ metric''. The velocity profile induced at infinity by the KMV$_0$ metric does not, however, resemble the one shown in Figure 1. In \cite{kn:shear} this geometry was generalized (by studying the Pleba\'nski--Demia\'nski family of metrics \cite{kn:plebdem,kn:grifpod}), to include both electric charge $Q$ and a parameter $\ell$ analogous\footnote{Analogous, but different: see \cite{kn:shear}. In the planar case, this kind of ``charge'', whose holographic interpretation will soon be given, does not give rise to the peculiarities of the usual Taub-NUT metric: see also \cite{kn:rehab}.} to NUT charge. These ``$\ell$QKMV$_0$ metrics'' do lead to the desired velocity profile, so we shall focus on them and related metrics in this work.

The $\ell$QKMV$_0$ metrics can be further generalized to incorporate magnetic charge: the \emph{dyonic} ``$\ell$dyKMV$_0$'' metrics so obtained are solutions of the AdS$_4$ Einstein-Maxwell equations (with cosmological constant $-3/L^2$) taking the form

\be
\label{A}
g(\ell {\rm dyKMV}_0)=-\frac{\Delta_r\Delta_\psi\rho^2}{\Sigma^2}\,\dif t^2+\frac{\rho^2\dif r^2}{\Delta_r}+\frac{\rho^2\dif\psi^2}{\Delta_\psi}+\frac{\Sigma^2}{\rho^2}\left[\omega\dif t-\dif\zeta\right]^2,
\ee
where
\ba
\label{B}
\rho^2&=&r^2+(\ell-a\psi)^2\cr
\Delta_r&=&\frac{(r^2+\ell^2)^2}{L^2}-8\pi M^*r+a^2+4\pi \left[Q^{*2}+P^{*2}\right]\cr
\Delta_\psi&=&1+\frac{\psi^2}{L^2}(2\ell-a\psi)^2\cr
\Sigma^2&=&(r^2+\ell^2)^2\Delta_\psi-\psi^2(2\ell-a\psi)^2\Delta_r\cr
\omega&=&\frac{\Delta_r\psi(2\ell-a\psi)-a(r^2+\ell^2)\Delta_\psi}{\Sigma^2}.
\ea
The corresponding electromagnetic potential one-form outside the black hole is
\begin{eqnarray}\label{C}
& &A(\ell {\rm dyKMV}_0)=\left[-\,\frac {Q^*r + P^*(\ell-a\psi)} {\rho^2L}+\frac{Q^*r_h+P^*\sqrt{\ell^2+aL}} {L(r_h^2+\ell^2+aL)}\right]\m{d}t \nonumber \\
& &  + \left[\frac{-Q^*r(2\ell-a\psi)\psi+P^*\left(\psi-\frac{\ell}{a}\right)(r^2+\ell^2)} {\rho^2L}-\frac{Q^*r_hL-P^*\frac{\sqrt{\ell^2+aL}}{a}(r_h^2+\ell^2)}{L(r_h^2+\ell^2+aL)}\right]\m{d}\zeta,
\end{eqnarray}
where the gauge has been fixed so that the Euclidean version is well-defined; see Appendix A for this procedure. Notice particularly that, in the presence of angular momentum, the timelike component depends on both $Q^*$ and $P^*$.

We now briefly summarize the salient properties of this black hole; for more detail, we refer the reader to Appendix B.

The coordinates are as follows: $t$ can be regarded as proper time at infinity (this fixes the conformal gauge), $r$ is a ``radial'' coordinate, and $\psi$ and $\zeta$ are dimensionless bulk ``angular'' coordinates\footnote{If $a$ and $\ell$ were zero, then one could impose periodic identifications on these coordinates, and then they would literally be angular coordinates (on a 2-torus). We do not do this here, however, because it leads to several problems, including the creation of closed timelike worldlines: see the corrigendum to \cite{kn:klemm}. Instead we simply confine attention to a finite domain in the ($\psi, \zeta$) plane.}, analogous to the spherical polar coordinates $\theta, \phi$. At infinity, they define the reaction plane coordinates: $\zeta$ is related to the collision axis coordinate $z$, and $\psi$ to the transverse coordinate $x$, through the equations
\begin{equation}\label{CC}
\m{d}x^2 = {\m{d}\psi^2L^2 \over 1+\frac{\psi^2}{L^2}(2\ell-a\psi)^2}, \;\;\; z = \zeta L.
\end{equation}
That is, with these relations, the spatial metric at infinity is $\m{d}x^2 + \m{d}z^2$ (because this metric is just ${\m{d}\psi^2L^2\over \Delta_{\psi}} + \m{d}\zeta^2L^2$ in the original coordinates).

The fact that this black hole is endowed with angular momentum is evidently related to the presence of the quantity $\omega$ defined in (\ref{B}) (since it is responsible for the off-diagonal component of the metric); and it is non-zero because of the parameters $a$ and $\ell$. The first has a clear interpretation: it is the (shearing) angular momentum per unit energy, or \emph{specific angular momentum}, of the black hole. (This can be shown either by adapting the methods explained in \cite{kn:klemm}, or by a simple heuristic procedure explained in Appendix B.) The physical interpretation of $\ell$ is less clear, but it can be revealed by the following argument.

The function $\omega(r,\psi)$ is the ``angular velocity''; it measures the effects of \emph{frame dragging}\footnote{If the event horizon were compactified (so that $\psi$ and $\zeta$ would be truly angular coordinates on a torus, as discussed above) then $\omega$ would indeed be an angular velocity, in the $\zeta$ direction. In fact, an object released in this spacetime with zero momentum in the $\zeta$ direction will be frame-dragged at a rate (as seen from infinity) given precisely by $\omega$: that is, $\m{d}\zeta/\m{d}t = \omega(r,\psi)$.}. A key point is that $\omega(r,\psi)$ is \emph{not} a constant at infinity ($r \rightarrow \infty$); instead it still depends on the remaining coordinate $\psi$: we have
\begin{equation}\label{D}
\omega_{\infty} = \psi (2\ell - a\psi)/L^2.
\end{equation}
In short, this black hole induces frame dragging at arbitrarily large distances from the event horizon. The form of $\omega_{\infty}$ as a function of $\psi$ is determined by both $\ell$ and $a$; in particular, the maximal frame-dragging effect occurs when $\psi = \ell/a$, and its magnitude is $\omega_{\infty}^{\m{max}}=\ell^2/aL^2$. This gives some insight as to the geometric meaning of $\ell$: for a given value of $a$, it controls the extent of frame-dragging far from the event horizon.

Now that we understand the black hole parameters $a$ and $\ell$, we can proceed to explain the meanings of $M^*$, $Q^*$, and $P^*$. These can in fact be defined by dividing the physical mass $M$, and the physical charges $Q$ and $P$, by a certain constant $\Gamma$, proportional to the area of the region of interest at infinity (see \cite{kn:peldan} and Appendix B of the present work). One could try to evaluate $\Gamma$ by studying the actual (finite) domain occupied by the plasma in the $(x, z)$ plane at some given time, but this is complicated, since the size of this domain varies from collision to collision in a way that is hard to parametrize. Instead, we proceed as follows.

Let $r = r_h$ at the event horizon. Then the \emph{mass per unit horizon area} $\mathcal{M}$, the electric charge per unit horizon area $\mathcal{Q}$, and the magnetic charge per unit horizon area $\mathcal{P}$, are given by
\begin{equation}\label{G}
\mathcal{M} \;=\;M^*K_V/(r^2_h+\ell^2),\;\;\;\;\mathcal{Q}\;=\;Q^*K_V/(r^2_h+\ell^2), \;\;\;\;\mathcal{P}\;=\;P^*K_V/(r^2_h+\ell^2),
\end{equation}
where $K_V$ is a dimensionless function of $V$ \emph{only}, defined by
\begin{equation}\label{GG}
K_V\;=\;\int_0^1{\m{d}p\over \sqrt{1+V^2p^2(2-p)^2}}.
\end{equation}
The derivations will be found in Appendix B below. (The point is that $K_V$ depends only on a quantity that we already understand, namely the maximal value of the velocity profile function, not on further details of the collision geometry.) We then take the view that $M^*$, $Q^*$, and $P^*$ are purely geometric parameters describing the bulk spacetime, while $\mathcal{M}, \mathcal{Q}$, and $\mathcal{P}$ are more directly physical; in particular, $\mathcal{M}$ has a direct holographic interpretation, as will be explained in the next section.

The entropy $S$ of the black hole can likewise be used to define $S^* \equiv S/\Gamma$, and then the entropy per unit horizon area is $S^*K_V/(r^2_h+\ell^2).$ But by Hawking's law of black hole entropy, this quantity is just one quarter in natural units, so we have the useful relation
\begin{equation}\label{H}
r^2_h+\ell^2\;=\;4S^*K_V.
\end{equation}

The Hawking temperature of this black hole can be computed most easily using the following device:
\begin{equation}\label{I}
T \;=\;{\m{d}M\over \m{d}S} \;=\; {\m{d}M^*\over \m{d}S^*},
\end{equation}
with the understanding that all other parameters are kept fixed. Using equation (\ref{H}), one can express the definition of $r_h$, $\Delta_r (r_h) = 0$, in the form
\begin{equation}\label{J}
\frac{16S^{*2}K_V^2}{L^2}-8\pi M^*\sqrt{4S^*K_V - \ell^2}+a^2+4\pi \left[Q^{*2}+P^{*2}\right] = 0.
\end{equation}
Differentiating this with respect to $S^*$ and applying equation (\ref{I}), one finds that
\begin{equation}\label{K}
T\;=\;{K_V(r_h^2 + \ell^2)\over \pi r_hL^2}\;-\;{2K_VM^*\over r_h^2}.
\end{equation}

We are now in a position to complete the holographic dictionary.

\addtocounter{section}{1}
\section* {\large{\textsf{3. The Dictionary Including Shear and a Magnetic Field}}}
All of the black hole parameters and functions described in the preceding section have non-trivial consequences for the physics at the conformal boundary; that is, they have \emph{holographic} interpretations. We now explain these.

We saw that the frame-dragging induced by the black hole angular momentum persists to infinity. This frame-dragging at infinity is responsible for the shearing effect portrayed in Figure 1; the velocity $v$ is just $\omega_{\infty}L$. Clearly we must confine $\psi$ to lie between $0$ and $\Psi \equiv \ell/a$, since that is the value of $\psi$ where the frame-dragging attains its maximum value; at that point, we have the maximal dimensionless velocity (equal to 0.4 in Figure 1), which we denoted earlier by $V$. So we have
\begin{equation}\label{E}
V\;=\;\omega_{\infty}^{\m{max}}L\;=\;\ell^2/aL.
\end{equation}
Put differently, we now see the physical (holographic) interpretation of $\ell$: given $V$ and $L$ (which has a natural interpretation in terms of the typical size of the collision zone), $\ell$ is just another way of representing the specific angular momentum\footnote{Consequently, it does not make sense to try to take $\ell$ to zero without simultaneously taking $a$ to zero.}: we have $\ell = \sqrt{VaL}$.

We should note in passing that while the velocity profile of the plasma has a very simple quadratic form (equation (\ref{D})) as a function of $\psi$, it is less simple as a function of the physical coordinate $x$: in fact we have
\begin{equation}\label{F}
v(x) = {V\over 3} - \wp\bigg({\sqrt{|a|}\over L^{3/2}}\big(x+\varepsilon\big);\; -4\bigg[1-{V^2\over 3}\bigg],\; {-8V \over 3}\bigg[1+{V^2\over 9}\bigg]\bigg),
\end{equation}
where on the right we have the Weierstrass $\wp$-function with elliptic parameters fixed by $V$. (Here $\varepsilon$ is a constant chosen so that the graph passes through the origin.) With certain choices of $V$, $a$, and $L$, this is the function shown in Figure 1.

The entropy per unit mass of the $\ell$dyKMV$_0$ black hole is 1/($4\mathcal{M}$), where $\mathcal{M}$ is given by one of the equations in (\ref{G}) above. Thus we have, using equation (\ref{E}),
\begin{equation}\label{L}
\varsigma_S\;=\;{1\over 4M^*\,K_V}\left(r_h^2 + VaL\right).
\end{equation}
This quantity will be interpreted holographically as the average entropy per unit energy of the newly formed QGP produced by heavy ion collisions in a given beam.

Notice that if $\varsigma_S$ is given, it follows from (\ref{L}) that $M^*$ should be regarded a function of $a$ (in a very complex way, because $r_h$ will also turn out to depend on $a$). Similarly we will see later that $Q^*$ and $P^*$ have to be functions of $a$. This is a little unusual, but there is no objection to it: we repeat that $M^*$, $Q^*$, and $P^*$ are \emph{geometric} parameters which we are free to vary in such a way as to produce desired behaviour in the \emph{physical} parameters. In other words, we are considering a particular curve, parametrised by $a$, in the space with coordinates ($M^*$, $Q^*$, $P^*$).

Again, using (\ref{E}), we can now express the Hawking temperature of the black hole as
\begin{equation}\label{M}
T\;=\;\left(1+{VaL\over r_h^2}\right)\left({K_V\,r_h\over \pi L^2}-{1\over 2\varsigma_S}\right).
\end{equation}
As usual, this will be interpreted holographically as the temperature of the QGP.

Understanding this function will be the central objective of this work. In order to proceed, however, we need to provide a holographic interpretation of one more quantity, namely $r_h$. That will be supplied by the holographic interpretation of the remaining black hole parameters $Q^*$ and $P^*$, to which we now turn.

The electromagnetic field 2-form at infinity does not vanish here: instead it is given (see equation (\ref{C})) by the simple expression
\begin{equation}\label{N}
F^{\infty}\;=\;{P^*\over L}\,\m{d}\psi \wedge \m{d}\zeta.
\end{equation}
Clearly the field at infinity is magnetic. In order to evaluate the field, we need to express $F^{\infty}$ in terms of a basis of forms which are of unit length relative to the metric at infinity. We saw earlier that the spatial component of this metric is ${\m{d}\psi^2L^2\over \Delta_{\psi}} + \m{d}\zeta^2L^2 = \m{d}x^2 + \m{d}z^2$ (see equations (\ref{CC})), so we have
\begin{equation}\label{O}
F^{\infty}\;=\;{P^*\sqrt{\Delta_{\psi}}\over L^3}\,\m{d}x \wedge \m{d}z\;=\;\,B^{\infty}(\psi)\,\m{d}x \wedge \m{d}z.
\end{equation}
Thus we see that the magnetic field at infinity is not a constant (as it is for the dyonic AdS black hole with a topologically spherical event horizon): it depends on $\psi$ (and also, through both $\Delta_{\psi}$ and $P^*$, on $a$). This is as it should be, since, as we know, the space at infinity here does not  ``move rigidly'': there is a non-trivial velocity profile, so one should expect the magnetic field likewise to be dependent on the transverse coordinate\footnote{We are not claiming that the specific dependence on the transverse coordinate implied by (\ref{O}) is realistic; only that the situation is not unreasonable qualitatively.}. In fact, the actual magnetic field does vary with transverse position; however, it varies slowly \cite{kn:ferrer}, and can be well approximated by its spatial mean. Similarly, denoting the spatial mean of $B^{\infty}(\psi)$ by $B_m$, we have
\begin{equation}\label{P}
B_m\;=\;{P^*\over L^3}\,J_V,
\end{equation}
where
\begin{equation}\label{Q}
J_V\;=\;{1\over \Psi}\int_0^{\Psi}\sqrt{\Delta_{\psi}}\m{d}\psi,
\end{equation}
where we recall again that $0 \leq \psi \leq \Psi = \ell/a = \sqrt{VL/a}.$ The quantity $J_V$ is a (complicated) expression involving elliptic integrals; contrary to appearances, it depends \emph{only} on $V$: see Appendix B. We will interpret $B_m$ as the holographic version of the magnetic field discussed in Section 1.

Finally, we need a holographic expression for the baryonic chemical potential $\mu_B$. As usual, it will be given by (three times) the value at infinity of the timelike component of the potential one-form, so from equations (\ref{C}) and (\ref{E}) we have
\begin{equation}\label{R}
\mu_B/3\;=\;\frac{Q^*r_h+P^*\sqrt{\ell^2+aL}} {L(r_h^2+\ell^2+aL)}\;=\;\frac{Q^*r_h+P^*\sqrt{(1+V)aL}} {L(r_h^2+(1+V)aL)}.
\end{equation}
Given $\mu_B$, $r_h$, $a$, and $P^*$, we can now compute $Q^*$. This supplies the holographic interpretation of $Q^*$: in essence, it is the bulk counterpart of the chemical potential, though in a more indirect sense than is usual.

This completes the holographic dictionary for our purposes: we have holographic interpretations for all of the parameters in the $\ell$dyKMV$_0$ geometry. We can now try to apply this dictionary to inverse catalysis.

\addtocounter{section}{1}
\section* {\large{\textsf{4. Inverse Magnetic/Shear Catalysis at $\mu_B = 0$}}}
At the RHIC, the collisions (apart from those in the beam energy scan programme) give rise to a plasma with a high antiparticle/particle ratio, and $\mu_B$ is therefore quite small, about an order of magnitude smaller than the temperature; that is, the specific baryonic chemical potential $\varsigma_B \equiv \mu_B/T$ is a small dimensionless number. (At the LHC, in the ALICE (and other) experiments, it is still smaller.) For a first orientation, therefore, we will take $\varsigma_B = \mu_B = 0$; we will relax this assumption in the next section. Notice from equation (\ref{R}) that, in order to have a vanishing $\mu_B$, we must ensure that the electric charge on the black hole is \emph{not} zero. This is in very sharp contrast to all previous holographic discussions of $\mu_B$; it is of course due to the presence of angular momentum.

We can now show how to compute $r_h$, the location of the event horizon. Its defining equation, $\Delta_r(r_h) = 0$, can be expressed as
\begin{equation}\label{S}
\frac{(r_h^2+VaL)^2}{L^2}- {2\pi (r_h^2+VaL)r_h\over \varsigma_S K_V}+a^2+{4\pi B_m^2L^6\over J_V^2} \left[1+{(1+V)aL\over r_h^2}\right]\;=\;0.
\end{equation}
Here we have used (\ref{E}) to express $\ell$ in terms of $a$ and $V$, (\ref{L}) to express $M^*$ in terms of $\varsigma_S$, $a$, and $V$, (\ref{R}) (with $\mu_B = 0$) to express $Q^*$ in terms of $P^*$, $a$, and $V$, and finally (\ref{P}) to express $P^*$ in terms of $B_m$ and $a$.

The final position is as follows. Specify $(a, B_m, V, \varsigma_S)$: all of these quantities have direct physical interpretations and can be considered, in principle at least, to be known\footnote{We also have to specify $L$. This can be computed from the geometry of the collision zone; we shall use $L = 10$ fm. For the details, see \cite{kn:shear}. Alternatively, one might take the view that $L$ should simply correspond to the transverse size of the system in question, and then 10 fm is again a reasonable estimate.}. Then (after computing $J_V$ and $K_V$) we can use equation (\ref{S}) to compute $r_h$. Finally, equation (\ref{M}) will allow us to compute the temperature, as predicted by holography.

Again, \emph{in principle} one should proceed as follows. Our starting point was that the magnetic field in this case is associated with the same processes as those responsible for the presence of a large specific angular momentum: that is, with the internal motions of the plasma as it arises from a peripheral collision of heavy ions. Therefore, we must regard $B_m$ as a function of $a$. For example, in the limiting case of a truly central collision, we would expect both the magnetic field and $a$ to vanish. (It follows from equation (\ref{P}) that $P^*$ must depend on $a$, and similarly for $Q^*$, as we predicted earlier.) The precise dependence of $B_m$ on $a$ is hard to specify, for reasons to be discussed; let us suppose for simplicity that the relation is linear:
\begin{equation}\label{T}
B_m\;=\;{\beth \,a\over L^3},
\end{equation}
where $\beth$ is a positive dimensionless constant. This is actually reasonable if one studies the time evolution of the plasma. A useful way to describe this evolution is in terms of a scale factor describing the expansion after the formation of the plasma: in many ways, this expansion is analogous to the expansion of the cosmic QGP after reheating \cite{kn:mocsy,kn:heinz,kn:raf}. Because both the specific angular momentum $a$ and the magnetic field $B_m$ are governed by  conservation laws (of angular momentum and magnetic flux, respectively), one expects them to evolve in much the same way, that is, with the inverse square of the scale factor. A relation like (\ref{T}) is therefore not unreasonable\footnote{We should stress however that (\ref{T}) is not essential: any reasonable values of $B_m$ and $a$ can be substituted into equation (\ref{S}).}.

Assuming (\ref{T}), then, we have
\begin{equation}\label{U}
\frac{(r_h^2+VaL)^2}{L^2}- {2\pi (r_h^2+VaL)r_h\over \varsigma_S K_V}+a^2+{4\pi \beth^2a^2\over J_V^2} \left[1+{(1+V)aL\over r_h^2}\right]\;=\;0.
\end{equation}
We can use this equation to regard $r_h$ as a function of a single variable, and then $T$ can likewise be regarded as a function of a single variable, $a$.

The amount of angular momentum transferred to the plasma depends on the collision geometry: in particular, it depends on the impact parameter. It is negligible both for very small and for very large impact parameters, and reaches a maximum in a certain optimal geometry \cite{kn:bec}. For a \emph{given beam}, then, we can think of $a$ as taking a range of values, according to the impact parameter, from zero up to some maximum. The objective is to study the corresponding variation of the critical temperature, using equations (\ref{M}) and (\ref{U}).

Unfortunately, at present it is not very reasonable to proceed in this idealized manner, because, for reasons to be explained, it is hard to estimate the maximal value of $a$. In any case, the gauge-gravity duality has not yet reached a point where precise predictions are possible. We need to insert some data in order to discern the \emph{trends} ---$\,$ in particular, we wish to know whether angular momentum reinforces or reduces the effects of magnetism ---$\,$ but we do \emph{not} claim that the actual numbers produced are to be taken as predicted values. Of course, we do however need to use quasi-realistic input data, so that the trends we find apply to the physical domain. In particular, we are interested in the vicinity of the quasi-critical temperature explored by the RHIC experiment, and to the vicinity of the critical end point in the quark matter phase diagram being investigated in current and near-future experiments (see the next section); so we use data relevant to those experiments. (Thus for example we do not consider magnetic fields in the GeV$^2$ range, interesting though those may be for other purposes.)

In view of all this, we will proceed more cautiously, as follows.

We begin by considering the plasma near to the (pseudo)-critical temperature $T_c$, which locates the crossover \cite{kn:aoki} between the plasma and hadronic states. At first we ignore the effects of magnetic/shear catalysis: that is, we temporarily set $a = B_m = 0$. Let us call this $T_{c0}$; one can think of it as the critical temperature for the QGP arising from \emph{central} collisions in a given beam. Recent lattice computations have determined the value of this parameter (when $\mu_B = 0$) with very considerable precision: in \cite{kn:bhat} one finds $T_{c0} = 155(1)(8)$ MeV (statistical)(systematic); for our purposes $T_{c0} \approx 0.75$ fm$^{-1}$ will however suffice. In this simplified holographic picture, this puts the specific entropy $\varsigma_S$ at around 2 fm (about 10 per baryon), a not unreasonable estimate. We also fix $V \approx 1$, because the magnetic field is only very intense for a short time \cite{kn:tuchin} after equilibration, and so $V$, the maximal velocity of the plasma, will be approximately the dimensionless velocity of either projectile, that is, essentially unity\footnote{This fixes the quantities $K_V$ and $J_V$: we have $K_1 \approx 0.82473$ and $J_1 \approx 1.22991$.}. Next, we ``turn on'' the magnetic field. For this purpose we use the maximal value estimated in \cite{kn:skokov} for peripheral collisions in the case of the RHIC experiment: that is, we take $B_m$ to be around $m_{\pi}^2$, where $m_{\pi}$ is the mass of the pion. We will use $(0.75$ fm$^{-1})^2 \approx 0.56$ fm$^{-2},$ but also some lower and higher values in order to discern a trend.

Next, we should consider the value of $a$ that corresponds to this choice of $B_m$. Unfortunately there are in fact several problems in estimating $a$. First, it is very sensitive to the details of the collision geometry, and also to the energy density of the collision. For example, in passing from the RHIC to the LHC, typical energy densities increase by a factor of 2 or 3, but \cite{kn:bec} the angular momentum increases by nearly two orders of magnitude. Since $a$ is the ratio of these two quantities, one sees that $a$ can increase quite dramatically from one physically reasonable situation to another. Second, one should note that the angular momentum transferred to the plasma is maximized by a value of the impact parameter which \emph{reduces} the energy density (compared to that of a central collision) to an extent that is hard to quantify. In \cite{kn:bec}, the angular momentum (which is dimensionless in these units) is estimated to peak at around $7.2 \times 10^4$ in RHIC collisions; a simplified computation of the overlap volume then leads to an angular momentum density of roughly 350 - 400 fm$^{-3}$. The usually cited value of the energy density for \emph{central} collisions, around 15 fm$^{-4}$, would then lead to an estimate of a maximal $a \approx 20-30 $ fm, but this would almost certainly be an underestimate; values in the range of $\approx 75$ fm may be reasonable, but much larger values may be possible.

Rather than try to guess a specific maximal value for $a$, then, we will use the data to select a reasonably wide range of possible values, and then use all of the above, together with equation (\ref{S}), to investigate the consequences for $T_c$.

Before we proceed, we should clarify one rather confusing point. When one studies (for example) the Kerr-Newman black hole, one finds that $r_h$ decreases if the angular momentum parameter (or the electric or magnetic charge) is increased, \emph{if} all of the other parameters are held constant. This allows one to predict the behaviour of the Hawking temperature under these circumstances. Here, however, the situation is very different: a glance at equations (\ref{S}) and (\ref{U}) shows that the parameter $a$ occurs in every term. That is, as we have emphasised repeatedly, the parameters $M^*, P^*$, and $Q^*$ are certainly \emph{not} being held constant as we vary $a$. We are therefore not entitled to expect that $r_h$ will necessarily become smaller as we increase $a$. But let us suppose that, nevertheless, it does ---$\,$ it turns out that this is the case when $\mu_B = 0$ (though not always when $\mu_B \neq 0$). Then equation (\ref{M}) still does not allow us to draw any conclusions regarding $T$. For while the second bracketed expression would then become smaller with increasing $a$, the first bracketed expression would become larger. In short, it is far from obvious whether $T$ will increase or decrease with $a$ in this situation; it is \emph{not} clear that holography predicts inverse magnetic/shear catalysis.

We summarize our results in the following table. Values of $T_{c}$ are given in fm$^{-1}$, of $a$ in fm, of $B_m$ in fm$^{-2}$. We have used the maximal value for $B_m$ suggested in \cite{kn:skokov}, which we accordingly denote by $B_{SIT}$, in the middle row, but also some somewhat lower and higher values (zero, 0.5, 1.5 and two times this value) for comparison. Bear in mind that we are taking $T_{c0} \approx 0.75$ fm$^{-1}$ here, when there is no magnetic field and no angular momentum (that is, in the corresponding central collisions).
\begin{center}
\begin{tabular}{|c|c|c|c|c|c|c|c|c|}
  \hline
 & $a$ = 0  & 25 & 50 &75 & 100 &150 &250 &500 \\
\hline
$B_m= 0$ &  0.75000  & 0.74956 & 0.74912 & 0.74867 & 0.74820 & 0.74725&0.74523 &0.73943\\
$0.5 \times B_{SIT}$ &  0.74685  & 0.74638 & 0.74592 & 0.74544 & 0.74495 & 0.74394&0.74180 &0.73567\\
$B_{SIT}$ &  0.73702  & 0.73647 & 0.73592 & 0.73535 & 0.73478 & 0.73359&0.73106 &0.72381\\
$1.5 \times B_{SIT}$ &  0.71914 & 0.71842 & 0.71769 & 0.71695 &0.71618 & 0.71461&0.71126&0.70152\\
$2 \times B_{SIT}$ &  0.68985  & 0.68877  & 0.68766 & 0.68653 & 0.68536 &0.68295&0.67775&0.66200
 \\
\hline
\end{tabular}
\end{center}
Evidently both a pure magnetic field, and a magnetic field accompanied by a shearing angular momentum, tend to reduce the critical temperature: \emph{holography predicts inverse magnetic/shear catalysis.} Furthermore, the effect of angular momentum in this case is always to reinforce the effect of the magnetic field. However, at least for the quasi-realistic parameter values used here, the effect of angular momentum is smaller than that of magnetism.

The lattice computations of \cite{kn:bhat} imply that, for the present ---$\,$ though, in view of the rate of progress of lattice computations, this will change in the near future ---$\,$ statistical and/or systematic effects will only allow us to ascribe significance to differences of temperature of the order of at least $0.04 - 0.05$ fm$^{-1}$. Since the differences indicated by the table are of roughly that order of magnitude or smaller, we have a reminder of our warning that the actual numbers in the table should not be accepted as a concrete prediction of gauge-gravity duality: only the trends are to be trusted.

The real question raised by these results, however, is this: \emph{can one give a theoretical account of the statement that angular momentum reinforces inverse magnetic catalysis?} As we will now show, this question becomes still sharper when a non-zero baryonic chemical potential is considered.

\addtocounter{section}{1}
\section* {\large{\textsf{5. Inverse Magnetic/Shear Catalysis at $\mu_B \neq 0$}}}
The experimental study of the QGP at non-zero $\mu_B$ is of great interest, with the RHIC beam scan programme under way \cite{kn:BEAM}, and new facilities (SHINE, NICA, GSI/FAIR) under construction \cite{kn:shine,kn:nica,kn:fair}. The hope is that these facilities will map out the QCD phase diagram \cite{kn:mohanty,kn:satz}; in particular, a major objective is to locate the \emph{critical end point} \cite{kn:race}, where the crossover becomes a true phase transition.

The analogue of inverse magnetic catalysis in this case would be a shift in the location of the critical end point when magnetic fields and shear are taken into account: see for example \cite{kn:magdy} (in particular, Figure 8 of that work). Again, it is interesting to consider whether holography predicts any such effect. As we will see, shear has unexpected consequences in this case.

There are various estimates of the possible location of the critical end point in the phase diagram (before the effects of any kind of catalysis are taken into account; that is, in effect, for central collisions); see \cite{kn:karsch} for a recent discussion. For the sake of definiteness we choose a location $T_0^{CEP}$ = 145 MeV, $\mu_{B0}^{CEP}$ = 300 MeV, compatible with Figure 6 of \cite{kn:karsch}; here the zero subscripts remind us that $a$ and $B_m$ have effectively been set to zero. The objective now is to study how this point might move under the influence of intense magnetic fields and large shearing angular momenta.

The procedure is much as before, though it is technically a little more involved. As before, we use the ``initial conditions'' at the point $\left(\mu_{B0}^{CEP}, T_0^{CEP}\right)$ to determine the coefficients in an equation that fixes the location of the event horizon in the bulk, and then we can compute the temperature. It is clear that, with our assumed values, $\varsigma_B \approx 2$; for $\varsigma_S$ we proceed as follows. We note first that equation (\ref{S}) has to be replaced here, to account for $\mu_B \neq 0$; using equation (\ref{R}) and $\mu_B = \varsigma_B T$ one finds the following equation:
\begin{eqnarray}\label{V}
\frac{(r_h^2+VaL)^2}{L^2}&-&{2\pi (r_h^2+VaL)r_h\over \varsigma_S K_V} \;+\; a^2 \;+\; {4\pi B_m^2L^6\over J_V^2}  \nonumber \\
  && +\;{4\pi\left({1\over 3}\varsigma_B\,TL\left[r_h^2+(1+V)aL\right]-{B_mL^3\over J_V}\sqrt{(1+V)aL}\right)^2\over r_h^2} \;=\; 0.
\end{eqnarray}
Evaluating this at $a = B_m = 0,\; T = T_0^{CEP}$, we have
\begin{equation}\label{W}
{r_{h0}^2\over L^2}\; -\; {2\pi r_{h0}\over \varsigma_S K_V} \;+\; {4\pi \over 9}\varsigma_B^2(T_0^{CEP})^2L^2 \;=\; 0,
\end{equation}
where $r_{h0}$ is the value of $r_h$ for this case. From equation (\ref{M}) we also have
\begin{equation}\label{X}
T_0^{CEP}\;=\;{K_V\,r_{h0}\over \pi L^2}-{1\over 2\varsigma_S}.
\end{equation}
Given $T_0^{CEP}$ and $\varsigma_B$ we can now solve (\ref{W}) and (\ref{X}) for $r_{h0}$ and $\varsigma_S$.
Setting $\varsigma_B \approx 2$, and $T_0^{CEP} \approx 0.7$ fm$^{-1}$ we find $\varsigma_S \approx 1.687$ fm.

Now we return to the general case, with $B_m \neq 0$ and $a \neq 0$. In order to make a comparison with the $\mu_B = 0$ case, we make use of the same values of $B_m$ and $a$ as in the preceding section. We substitute all of these quantities into the full versions of equations (\ref{M}) and ($\ref{V}$), solving them simultaneously for the two remaining unknowns, $r_h$ and $T = T^{CEP}$. (There are two real sets of solutions, corresponding to two horizons, inner and outer; we are interested only in the outer horizon, so we choose the solution for $T$ corresponding to the larger of the solutions for $r_h$.) The results for $T^{CEP}$ are shown in the table; bear in mind that the comparison should be with $T_0^{CEP} \approx 0.7$ fm$^{-1}$.

\begin{center}
\begin{tabular}{|c|c|c|c|c|c|c|c|c|}
  \hline
 & $a$ = 0  &  25 & 50 & 75 & 100 & 150 &250 &500 \\
\hline
$B_m = 0$ &  0.70000  & 0.69871 & 0.69741 & 0.69609 & 0.69475 & 0.69203&0.68637 &0.67088\\
$0.5 \times B_{SIT}$ &  0.69765  & 0.69850 & 0.69809 & 0.69747 & 0.69673 & 0.69503&0.69111 &0.67935\\
$B_{SIT}$ &  0.69038  & 0.69346 & 0.69393 & 0.69399 & 0.69382 & 0.69309&0.69072 &0.68209\\
$1.5 \times B_{SIT}$ & 0.67748 & 0.68304 & 0.68447 & 0.68523 &0.68564 & 0.68587&0.68499&0.67912\\
$2 \times B_{SIT}$ &  0.65734  & 0.66600  & 0.66854 & 0.67010 & 0.67116 &0.67242&0.67307&0.66976
 \\
\hline
\end{tabular}
\end{center}
We see at once that the presence of a non-zero baryonic chemical potential has dramatic consequences.

$\bullet$ All other values in the table are smaller than the value for $B_m = a = 0$; \emph{the effects of a combined magnetic field/shear are always in the direction of inverse catalysis}\footnote{We have checked much larger (and smaller) values of $a$ and $B_m$, and always find the same result (as in \cite{kn:gergely}).}.

$\bullet$ If we ignore angular momentum, then we will (\emph{wrongly}) conclude that the magnetic field necessarily has a straightforward effect on the temperature of the critical end point: the stronger the field, the lower the temperature. (Equally, if we ignore the magnetic field, we will conclude wrongly that angular momentum has a similarly straightforward, though less marked, effect.)

$\bullet$ However, if we take both effects into account, \emph{it is no longer the case} that increasing the magnetic field, or increasing the angular momentum, \emph{always} tends to lower the temperature. In both cases, fixing one (at some not very small value) and regarding $T^{CEP}$ as a function of the other, we find that it rises to a maximum (always still lower than $T_0^{CEP}$) before resuming the expected decline at large values of the parameter being varied. For example, for $B_m = B_{SIT}$, $T^{CEP}$ rises as $a$ increases from zero to around 75 fm; only after that does it begin to decrease. (Note that, according to our earlier discussion, $a = 75$ fm may well be a realistic maximal estimate corresponding to this value of the magnetic field.) For larger values of $B_m$, the maximum tends to occur at larger values of $a$. On the other hand, if we fix $a$ at 75 fm, and increase $B_m$ from zero, one again finds that the temperature reaches a maximum somewhat below $B_m = B_{SIT}$. In short, there is a ``ridge'' in the graph of $T^{CEP}$ as a function on the $(a,\,B_m)$ plane.

$\bullet$ As we have stressed, in reality $a$ and $B_m$ cannot be varied independently: they are inextricably combined. If one is considering collisions, in a given beam, at various impact parameters, one should therefore fix a relationship between them, and then regard $T^{CEP}$ as a function along the corresponding line in the $(a,\,B_m)$ plane. This line may or may not encounter the ``ridge''. For example, let us adopt the simplest possible relation, given by equation (\ref{T}) above, and assume for the sake of argument that the line runs through the point $(a = 75$ fm, $B_m = B_{SIT})$. Then one finds that this line does not meet the ``ridge'', and the temperature decreases monotonically along it as one moves away from the origin. Even in this case, however, the decline is less steep than it would be if one neglected angular momentum: the ``ridge'' makes its presence felt indirectly.

In short: holography continues to predict inverse magnetic/shear catalysis when $\mu_B \neq 0$; but it also indicates that the dependence of the critical end point temperature on $B_m$ and $a$ will not be simple, and that ignoring one or the other will lead to a potentially serious oversimplification. In particular, one may well find that neglecting angular momentum will result in an \emph{over-estimate of the ability of magnetic fields to shift the location of the critical end point.}

\addtocounter{section}{1}
\section* {\large{\textsf{6. Conclusion}}}
Inverse catalysis is a remarkable and unexpected phenomenon, and, in view of the current intense interest in establishing a firm basis for the quark matter phase diagram, it is potentially a matter of direct experimental concern. However, its theoretical basis is also extremely important, and much effort is currently being devoted to this very question. These theories involve many subtle concepts fundamental to our understanding of QCD \cite{kn:bruck,kn:krein}.

Any theory of this phenomenon will need to be subjected to tests, and we propose that the influence of angular momentum, clearly indicated by holography, might provide such a test. We have argued that such a theory should be able to account for two phenomena: a simple enhancement of the effect of the magnetic field by shear when the baryonic chemical potential $\mu_B$ is small, and a much more complex pattern, involving a ``ridge'' in the $(a,\,B_m)$ plane, when $\mu_B \neq 0$.

The physical origin of the ``ridge'' clearly merits a more detailed investigation. It may, of course, merely be an indication that the holographic description fails under these circumstances. It is probably more useful, however, to regard these results as a hint that the QGP may behave in an unexpected way in these extreme conditions, with large values of the net baryon density, large magnetic fields, and a strong shearing effect. For example, it is well known that, in some systems, there is a useful analogy between rotation and magnetic fields: this is a major theme in the theory of the quantum Hall effect \cite{kn:viefers}, and attempts have been made to extend the analogy to the relativistic case \cite{kn:mameda}. More recently, it has been proposed that there is a similar analogy between density (essentially, the baryonic chemical potential) and rotation \cite{kn:fuku}. Our results here suggest that, when the QGP is \emph{sheared rather than rotated}, these analogies break down in some way. It may be that an understanding of the ``ridge'' can be gained by extending the discussions in \cite{kn:mameda} and \cite{kn:fuku} to deal with shear instead of rotation.

The first step would be to determine whether the ``ridge'' is still present in a holographic description of the \emph{rotating} QGP ---$\,$ recall that the QGP might indeed rotate in some circumstances (for example, if the viscosity is large): see \cite{kn:KelvinHelm,kn:viscous,kn:csernairecent1,kn:csernairecent2,kn:nagy}. If it is not present in that case, this would be an indication that \emph{some specific aspect of the shearing motion itself} is responsible for the breakdown of the magnetic/rotation or the density/rotation analogies in these conditions. Our preliminary investigations suggest that the holography of a rotating plasma differs from the shearing case to a surprising degree; we hope to report results comparing the two situations in the near future.

\addtocounter{section}{1}
\section*{\large{\textsf{Acknowledgements}}}
The author wishes to acknowledge helpful discussions with Prof Soon Wanmei, and with Jude and Cate McInnes.

\addtocounter{section}{1}
\section*{\large{\textsf{Appendix A: The Electromagnetic Potential Form}}}
The general form of the electromagnetic one-form is found by a straightforward extension of the methods given in \cite{kn:dyon}. The problem is to determine the gauge-fixing constant terms in equation (\ref{C}) above.

We begin with the general expression
\begin{eqnarray}\label{COG}
& &A(\ell {\rm dyKMV}_0)=\left[-\,\frac {Q^*r + P^*(\ell-a\psi)} {\rho^2L}+\kappa_t\right]\m{d}t \nonumber \\
& &  + \left[\frac{-Q^*r(2\ell-a\psi)\psi+P^*\left(\psi-\frac{\ell}{a}\right)(r^2+\ell^2)} {\rho^2L}+\kappa_{\zeta}\right]\m{d}\zeta,
\end{eqnarray}
where $\kappa_t$ and $\kappa_{\zeta}$ are constants to be determined. The simplest way to understand why these constants must be present is to switch to the Euclidean version of the geometry. This is done by complexifying the coordinate $t$, together with $a$, $\ell$, $Q^*$, and $V$, (but \emph{not} $\psi$ and $P^*$). This procedure affects many of the quantities in the metric ---$\,$ for example, one has a Euclidean version of $\Delta_{\psi}$, namely $\Delta^E_{\psi} = 1-\frac{\psi^2}{L^2}(2\ell-a\psi)^2$, and similarly one has a function $\Delta_r^E$, which defines a ``Euclidean event horizon radius'' $r_h^E$ by $\Delta_r^E(r_h^E) = 0$.

Now consider the points in the Euclidean section with  $\psi = {\ell-\sqrt{\ell^2 + aL}\over a}$; at these points $\Delta_{\psi}^E = 0$ (giving rise to a coordinate ``singularity''). Next, at this value of $\psi$, take the limit $r \rightarrow r_h^E$. Then it is clear from the form of the metric that, at these points, the vectors $\partial_t$ and $\partial_{\zeta}$ have zero norms with respect to the Euclidean version of the metric. Precisely because the geometry is Euclidean, it follows that the vectors themselves must vanish at these points, and consequently we must have $A(\ell {\rm dyKMV}_0)^E(\partial_t) = A(\ell {\rm dyKMV}_0)^E(\partial_{\zeta}) = 0$ there if the potential is to be non-singular; here $A(\ell {\rm dyKMV}_0)^E$ is the Euclidean version of the potential one-form. Substituting $\psi = {\ell-\sqrt{\ell^2 + aL}\over a}$ and $r = r_h^E$ into the Euclidean version of equation (\ref{COG}), and returning to the Lorentzian section (taking care not to complexify any part of ${\ell-\sqrt{\ell^2 + aL}\over a}$, since it is just a particular value of the spacelike coordinate $\psi$), one obtains $\kappa_t = \frac{Q^*r_h+P^*\sqrt{\ell^2+aL}} {L(r_h^2+\ell^2+aL)}$, and similarly for $\kappa_{\zeta}$, and so we have equation (\ref{C}). (A more complete discussion of related questions may be found in \cite{kn:shear}, which uses the techniques of \cite{kn:chen}.)

\addtocounter{section}{1}
\section*{\large{\textsf{Appendix B: The Parameters $M^*, Q^*, P^*,$ and $a$}}}
The plasma occupies, at any given time, a finite domain in the $(x,z)$ plane; this domain is approximately rectangular, its dimensions depending on the size of the ions, the impact parameter, and the degree of relativistic contraction in the $z$ direction. This will define an approximately rectangular domain in the $(\psi, \zeta)$ plane. We will focus on the rectangle $[0, \Psi] \times [0, Z]$, where we recall that $\Psi = \ell/a = \sqrt{VL/a}$. The area of this domain (computed using the metric ${\m{d}\psi^2\over \Delta_{\psi}} + \m{d}\zeta^2$) can be expressed as $K_V\Psi Z$, where
\begin{equation}\label{DAD}
K_V\;=\;{1\over \Psi}\int_0^{\Psi}{\m{d}\psi\over \sqrt{\Delta_{\psi}}} = \int_0^1{\m{d}p\over \sqrt{1+V^2p^2(2-p)^2}},
\end{equation}
where $p = \psi/\Psi = a\psi/\ell$. Note that the dependence on $a$ drops out. Similarly, the quantity $J_V$ defined by equation (\ref{Q}) can be expressed as
$J_V = \int_0^1\sqrt{1+V^2p^2(2-p)^2}\m{d}p,$ which likewise is independent of $a$. Both $K_V$ and $J_V$ are slowly varying functions of $V$: see Figures 2 and 3 respectively.

\begin{figure}[!h]
\centering
\includegraphics[width=0.55\textwidth]{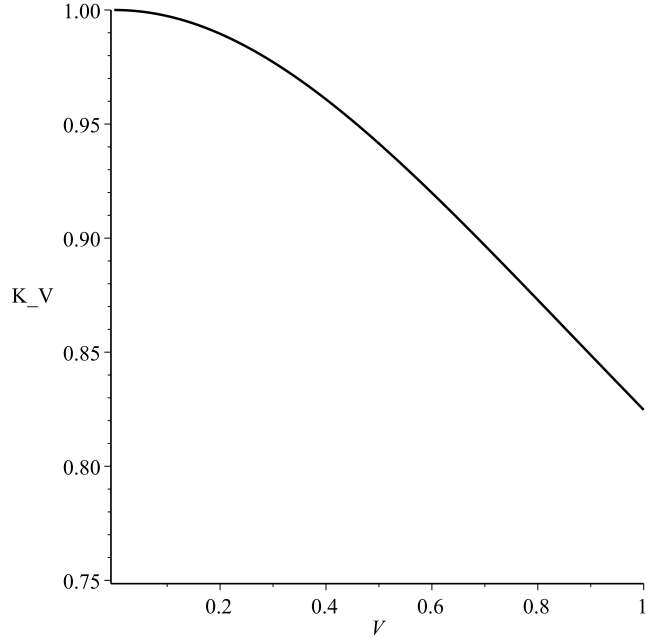}
\caption{$K_V$ for $0 \leq V \leq 1$.}
\end{figure}

\begin{figure}[!h]
\centering
\includegraphics[width=0.55\textwidth]{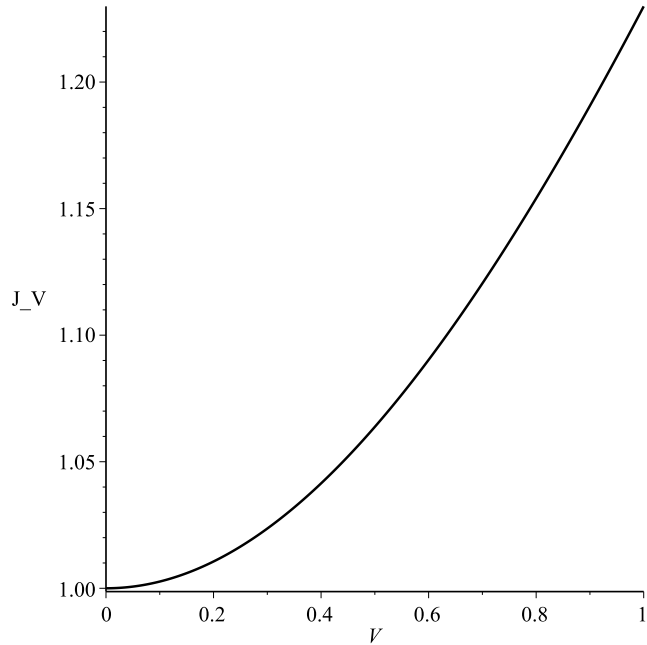}
\caption{$J_V$ for $0 \leq V \leq 1$.}
\end{figure}

We may now proceed to explain the meanings of the parameters $M^*, Q^*,$ and $P^*$. One can show \cite{kn:peldan} that the coefficient of the $1/r$ term in the black hole metric is given, up to a universal factor (equal to $-8\pi$ in four dimensions) by dividing the physical mass by the area of the ``angular'' part of the metric at conformal infinity. For example, for the AdS-Schwarzschild metric, this quantity is just $4\pi$, leading to the familiar $-\,2M$ coefficient. Here the corresponding quantity is $K_V\Psi Z$ and so the coefficient is $M^* = M/(K_V\Psi Z)$ and similarly for the electric and magnetic charge parameters $Q^* = Q/(K_V\Psi Z)$ and $P^* = P/(K_V\Psi Z)$.

Next, we consider the metric on the horizon, at fixed time: it is (from equations (\ref{A}) and (\ref{B}))
\begin{equation}\label{DOG}
h_{\psi, \zeta}\;=\;(r_h^2+\ell^2)\left[G(\psi)\m{d}\psi^2 + {\m{d}\zeta^2\over G(\psi)}\right],
\end{equation}
where
\begin{equation}\label{DOGGONE}
G(\psi)\;=\;{r_h^2+(\ell - a\psi)^2\over (r_h^2 + \ell^2)(1+\frac{\psi^2}{L^2}(2\ell-a\psi)^2)}.
\end{equation}
One sees from this that the \emph{mass per unit horizon area} is $M/[\Psi Z(r^2_h+\ell^2)] = M^*K_V/(r^2_h+\ell^2)$; similarly  $Q^*K_V/(r^2_h+\ell^2)$ and $P^*K_V/(r^2_h+\ell^2)$ are the electric and magnetic charges per unit horizon area. This explains the formulae (\ref{G}) for $\mathcal{M}, \mathcal{Q}$, and $\mathcal{P}$.

Finally, the specific angular momentum of any black hole can be computed, up to an overall sign, as $R^2\Omega_h$, where $R$ is the ``areal radius'' of the event horizon and $\Omega_h$ is its angular velocity (measured relative to the rotating ---$\,$ or, in our case, shearing ---$\,$ frame at infinity, as in \cite{kn:hawrot}, see also \cite{kn:gibperry}). For example, the areal radius of the AdS-Kerr black hole event horizon is given by $R^2 = (r_h^2+a^2)/(1-a^2/L^2),$ and the relevant angular velocity by $a(1-a^2/L^2)/(r_h^2+a^2)$, and so indeed $a = R^2\Omega_h$. In view of equation (\ref{DOG}), we have here $R^2 = r_h^2 + \ell^2,$ and from the final two members of (\ref{B}) we have $\Omega_h = -a/(r_h^2+\ell^2);$ so clearly $a$ is again to be interpreted as the specific angular momentum of the black hole.

\end{document}